\documentclass[12pt, a3paper]{article}

\usepackage{authblk}
\usepackage{longtable}
\usepackage{multirow}
\usepackage{hyperref}
\usepackage{array}
\usepackage{amsmath}
\usepackage{graphicx} 
\begin{document}
\title{
CloudSense: A Model for Cloud Type Identification using Machine Learning from Radar data
}
\author[1,2]{\small Mehzooz Nizar}
\author[1]{\small Jha K. Ambuj}
\author[1,3]{\small Manmeet Singh} 
\author[1]{\small Vaisakh S.B} 
\author[1]{\small G. Pandithurai}
\affil[1]{\footnotesize Indian Institute of Tropical Meteorology, Ministry of Earth Sciences, Pune, India}
\affil[2]{\footnotesize Cochin University of Science and Technology, Kochi, India}
\affil[3]{\footnotesize Jackson School of Geosciences, The University of Texas at Austin, Austin, Texas, USA}
\date{}

\maketitle
\textbf{ABSTRACT}
The knowledge of type of precipitating cloud is crucial for radar based quantitative estimates of precipitation. We propose a novel model called CloudSense which 
uses machine learning to accurately identify the type of precipitating clouds over the complex terrain locations in the Western Ghats (WGs) of India.  CloudSense 
uses vertical reflectivity profiles collected during July-August 2018 from an X-band radar to classify clouds into four categories namely stratiform,mixed 
stratiform-convective,convective and shallow clouds. The machine learning(ML) model used in CloudSense was trained using a dataset balanced by Synthetic 
Minority Oversampling Technique (SMOTE), with features selected based on physical characteristics relevant to different cloud types. Among various ML models 
evaluated Light Gradient Boosting Machine (LightGBM) demonstrate superior performance in classifying cloud types with a BAC of 0.8 and F1-Score of 0.82 . 
CloudSense generated results are also compared against conventional radar algorithms and we find that CloudSense performs better than radar algorithms. For 200 
samples tested, the radar algorithm achieved a BAC of 0.69 and F1-Score of 0.68, whereas CloudSense achieved a BAC and F1-Score of 0.77. Our results show that ML 
based approach can provide more accurate cloud detection and classification which would be useful to improve precipitation estimates over the complex terrain of 
the WG.\\

\textbf{Keywords: }Machine learning, Precipitating clouds, Doppler weather radar, Western Ghats, LightGBM

\section{INTRODUCTION}

Clouds are an integral component of convection and precipitation and play a vital role in modulating the global circulation, Earth’s radiative budget and 
hydrological cycle. Different cloud types are associated with different microphysical and radiative properties which influence the vertical distribution of 
heating in the atmosphere (Houze 1982; Houze 1997; Schumacher and Houze 2003). Improved knowledge of cloud types can improve weather and climate predictions 
through better representation of clouds in atmospheric models. From a regional perspective, accurate detection and classification of precipitating clouds are 
crucial for accurate quantitative precipitation estimation (QPE) which is widely used for extreme weather forecasting, climate studies and hydrological 
applications (Steiner and Houze, 1997; Gourley and Vieux 2005; Kühnlein et al.,2014; Thompson et al., 2015; Arulraj and Barros,2019). QPE algorithms provide rain 
estimates based on relationships specific to precipitating cloud type (Rao et al. 2001; Auipong and Trivej 2018) and therefore, any misclassification of 
precipitating clouds will impact the accuracy of QPE.

Earlier studies classified precipitating clouds either using space-borne active and passive sensors (Inoue 1987; Sassen and Wang 2008; Subrahmanyam and Kumar 
2013; So and Shin 2018) or ground-based radars and gauges (Penide et al., 2013; Loh et al., 2020; Zuo et al.,2022). Satellite-based classification though has 
wider spatial coverage but has limitations due to coarser spatial and temporal sampling. Broadly, precipitating clouds are categorized as convective and 
stratiform types (Houze 2014). Convective clouds are characterised by strong vertical air currents, small areal coverage, and high rainfall intensities while the 
converse is associated with stratiform clouds.

Most techniques apply a threshold value to distinguish between different precipitation types. For example, when precipitation intensity recorded by rain gauge 
exceeds a certain threshold, the precipitation cloud is considered convective but otherwise stratiform (Austin and Houze 1972; Houze 1973; Testud et al. 2001). 
Some studies used threshold-based techniques to raindrop spectra measured by surface disdrometer to classify precipitating systems (Waldvogel 1974; Tokay and 
Short 1996; Lavanya and Kirankumar 2021). However, these methods provide point measurements and can easily misinterpret the areal extent of precipitation in a 
convective cloud when there is weak precipitation intensity in its vicinity (Ran et al. 2021). Further, these methods rely only on rainfall measured near the 
surface and lack information on the vertical structure of clouds.

Ground-based weather radars are excellent tools for cloud and precipitation observations (Battan 1973; Fabry 2015). They provide 3D structures of clouds and rainfall at finer spatio-temporal scales which can be utilised for more accurate classification of precipitating systems (Xiao and Liu 2007). Ground-based radars are also used for calibration and validation of space-based precipitation products (Zhong et al. 2017; Biswas and Chandrashekhar 2018). A Doppler radar measures three base data parameters, viz., reflectivity, Doppler velocity and spectrum width. Using radar reflectivity threshold, Churchill and Houze (1984) identified the core of the convective precipitation and separated convective and stratiform regions within cloud clusters. They used a fixed radius of influence to identify the convective area of the core. The technique tends to overestimate the extent of the area designated as convective precipitation. Numerous echoes, despite retaining evident bright band characteristics, are categorized as convective. Steiner et al. (1995) modified this method using a variable radius of influence and used criteria of intensity, peakedness and surrounding area. However, the major limitation of Steiner’s algorithm was that it does not account for the vertical storm structure information which could improve the accuracy of precipitation classification. Later studies modified Steiner’s algorithm using additional information from the 3-D precipitation field and showed that the accuracy of the algorithm is improved (DeMott et al. 1995; Biggerstaff and Listemaa 2000). In contrast to studies which used only reflectivity, Williams et al. (1995) classified precipitating clouds through the synergy of vertical structures of reflectivity, Doppler velocity and spectrum width and illustrated the performance of their algorithm. However, the basic problem with conventional radar-based algorithms is sensitivity to the adopted threshold of a radar parameter. If the threshold reflectivity is set to be low, it is likely that stratiform regions are misclassified as convective. On the other hand, if the threshold is high, transitioning convective regions or mixed stratiform-convective areas may be identified as stratiform. Cloud structures are often complex which makes it difficult to distinguish them into different types. For example, a convective cloud in transition to stratiform exhibits mixed characteristics and these algorithms may fail in categorising them. Further, the threshold value is not always constant and needs to be adjusted and fine-tuned for different geographical locations, seasons and with changes in calibration constants of the radar.

In an attempt to solve this long-standing problem, studies have been conducted in the past that used machine learning (ML) models for classifying precipitating clouds into convective and stratiform (Anagnostou 2004; Lazri and Ameur 2018; Wang H et al., 2019; Yang et al., 2019; Wang Y et al., 2021; Ran et al., 2021; Ghada W et al., 2022). Wang H et al., 2019  developed a deep learning method of precipitation cloud type identification, based on the data of a dual-polarization Doppler weather radar. Anagnostou 2004 used a neural network approach to classify reflectivity measurements from ground-based volume scanning radar into convective and stratiform precipitation types. Ghada W et al., 2022, using a total of 20,979 min of rain data measured by a Micro Rain Radar (MRR) and built seven types of ML models for stratiform and convective rain type classification. Their results indicate that tree-based ensembles such as xgboost and random forest, performed better than simpler models. 

Over the Indian region, ML-based classification of precipitating clouds is either lacking or scarce. In this work, we develop a model using ML named as 
“CloudSense” and attempt to classify cloud types from radar reflectivity profiles in the Western Ghats (WG) of India.  This study primarily uses X-band radar 
data collected over a high altitude cloud physics laboratory (HACPL) at Mahabaleshwar (17.92° N, 73.66° E, $\sim$ 1.38 km MSL) in the WG (Fig 1). The radar site 
(referred to as MDV) is located at a distance of 26 km from HACPL and the data collected by the radar during July-August 2018 have been used for analysis.  The 
WG is one of the heaviest rainfall-receiving regions ($\sim$ 6000 mm) during the Indian summer monsoon(ISM) (Nandargi S, Mulye S 2012) and forms an ideal test bed for 
cloud and precipitation studies. The topography of the WG exerts an influence on monsoon rainfall in a manner that amplifies rainfall on the windward side and 
inhibits it on the leeward side. Due to complex topographical features, in-situ measurements of clouds and rainfall are scarce in this region. The rainfall in 
the WG is contributed by cloud systems such as stratiform, mixed stratiform-convective, convective and shallow convective clouds ( Konwar et al., 2012a; 
Maheskumar et al., 2014; Deshpande et al., 2015; Das et al., 2017).  Of all clouds, shallow convective clouds occur most frequently and contribute majorly to the 
total precipitation during ISM (Konwar et al., 2014). Earlier, Das et al., 2017 used MRR and X-band radar to study rain DSD characteristics for different cloud 
types over HACPL. 

In the present work, we use several ML models to classify cloud types from vertical profiles of reflectivity (VPRs) and select an optimal ML model to construct CloudSense. The model-generated results are compared against conventional radar algorithms and test cases are presented. The paper is organized as follows. Data and methodology are given in Sect. 2, and Sect. 3 contains results followed by conclusions in Sect. 4.

\section{DATA AND METHODOLOGY}

\subsection{Data}

The Indian Institute of Tropical Meteorology (IITM), Pune, has deployed a magnetron‐based mobile dual-polarized X-band Doppler radar at  Mandhardev (18.04°N, 73.86°E and $\sim$ 1.3km MSL), a remote location in the WG. The radar operates at $\sim$ 9.535 GHz with a peak power of about 200 kW. The technical speciﬁcations of the X-band radar are given in Table 1. The scan strategy of the radar consists of Range Height Indicator (RHI) and volume scan modes, repeated every 12 minutes. Each volume scan consists of plan position indicator (PPI) scans at 18 elevation angles from 0.5° to 90° and takes approximately 12 minutes.  On the other hand, the RHI scan is executed along 239° azimuth which points over the HACPL site (see Fig. 1).

The most important parameter provided by a weather radar is the equivalent reflectivity factor (Z, henceforth). Z is proportional to the 6th power of hydrometeor size in the scattering volume (Marshall et al. 1947) . The relation between Z and raindrop size is given by
\begin{equation}\label{1}
    Z= \int_{0}^{\infty}N(D) D^6 dD                                                                   
\end{equation}
where D is the raindrop diameter and N(D) is the precipitation particle size distribution.

In this work, Z data collected during July-August 2018 has been used. Several quality checks and corrections have been done for Z before analysis.  Mainly VPR data taken from RHI scans have been used in this study (Fig. 2a). Over HACPL, zenith pointing radar observations are lacking and therefore, vertical profiles of Doppler velocity (DV) and spectrum width (W) are not available. PPI data corresponding to 85º elevation can serve as pseudo zenith pointing observations and therefore, used here to obtain vertical profiles of   DV and W over MDV(Fig.2b).

\subsection{Methodology}

\subsubsection{Processing of Z data} 

Z data collected by the radar is in polar coordinates. Therefore, RHI data is converted to 2 D Cartesian coordinates with a grid spacing of 1 km in the horizontal and 100 m in the vertical. The data below 0.5 km are discarded to avoid noise and clutter near the ground. VPRs with Z$<$5 dBZ at 1.5 km above the ground are removed to avoid weak and noisy signals and are considered non-precipitating in nature. Vertical profiles of Z, DV and W are directly taken from 85º PPI data by averaging the respective profiles. Contoured frequency by altitude diagrams (CFAD) of radar parameters are also constructed (Yuter and Houze 1995). The bin size considered for CFADs is 5 dBZ for Z and 1 m/s for DV and W. The height interval is 0.1 km between 0.5 and 11 km. The number of occurrences at each level is standardized by the mean and standard deviation considering all reflectivity bins among all vertical levels.

\subsubsection{Labelling}

The precipitating clouds are classified into four categories: shallow, convective, stratiform and mixed stratiform-convective (also referred as mixed clouds) (Williams et al. 1995). Das et al. (2017) have classified clouds over HACPL into these four categories based on VPR data collected using MRR and the X-band radar. Here, we have adopted the VPR based definitions of cloud types described in their study. Note that here the cloud profiles are grouped not through a conventional approach based on radar algorithms rather they are manually labelled into these 4 categories after careful visual inspection of each VPR. Fig 3 shows the logical flow diagram of the classification algorithm. The first criterion is to check whether hydrometeors exist well above the melting layer. The climatological melting layer over the study region has been reported at 4 km (Kalapureddy et al. 2023); if the cloud top height is below 4 km, the cloud is classified as shallow otherwise the VPR is scrutinized for a bright band signature. The radar bright band is a prominent feature of stratiform precipitation and appears as a narrow region of enhanced Z just below the 0 C level in radar returns (Marshall et al. 1947; Battan 1973). If no bright band is detected, the cloud is placed in the convective category. Alternatively, if a bright band is present, further examination focuses on its strength. Mature stratiform clouds exhibit well-defined bright band features, while mixed clouds display embedded convection along with a less pronounced bright band (Das et al. 2017). Therefore, VPRs with strong and prominent bright band signatures are recognized as stratiform whereas those lacking this feature are labelled as mixed stratiform-convective clouds. These clouds being manually labelled are considered as true cloud types and the model results are tested against them.

\subsubsection{ CloudSense}

The CloudSense utilizes a ML model for the classification of defined cloud types except shallow clouds. Shallow clouds are not considered in the ML model algorithm due to the fact that they are generally confined below the melting layer (Fig 4d) which makes their identification $>$95 \% accurate. Shallow clouds are obtained by adopting a flexible height threshold of 4 km. For example, if a cloud crosses the 4 km melting layer but only by a marginal degree, it will be identified as shallow and therefore, no rigid threshold criterion is required. For other 3 cloud types, the ML model is trained using VPR samples and then tested for performance. Fig 4 shows the architecture of CloudSense.

The number of VPR samples for 3 cloud types is 1492 which includes 808 stratiform, 240 mixed and 444 convective clouds. These samples have been used for training the ML model. Stratiform clouds account for 54 \% of the total occurrences which is in agreement with the observations of Das et al. 2017. However, this leads to an imbalanced dataset. To address this, we use a Synthetic Minority Oversampling Technique (SMOTE)  (Chawla et al., 2002, Barua et al., 2011; Zhu et al., 2017; Elreedy and Atiya 2019) to make the dataset balanced by increasing the number of samples of convective and mixed clouds equal to the majority class(see supplementary Fig 1). Fig. 5 shows the CFADs of Z for all 4 cloud types segregated through manual labelling after oversampling.

Different cloud types have different physical processes which are linked to their vertical structures. For training a ML model, it is important to select features specific to each cloud type which are used as predictors for modelling. The performance of the ML model depends on these features and therefore, should be carefully selected. For identifying features, the following 3 zones are considered for the 3 types of clouds:

a) low-level (1-2 km): This zone is associated with rain drops and shows higher Z in case of convective clouds due to the presence of bigger rain drops. 

b) mid-level (3-5 km): This zone is associated with melting layer, has mixed phase hydrometeors and shows a prominent bright band peak for mature stratiform clouds. Case studies conducted over the WG have found bright band region situated between $\sim$3-5 km (Jha et al. 2019; Devisetty et al. 2019).

c) high-level (6-7 km): This zone is above the melting layer and mostly populated with ice phase particles. For mixed clouds, higher turbulence is observed in this zone (Williams et al. 1995); however, Z does not show any contrasting features.

A total of 41 predictors are considered from the defined 3 zones. 

\subsubsection{Tuning and Evaluating the model}

ML algorithms have several hyperparameters which determine the overall performance of the algorithm. Different hyperparameter values produce different model parameter values for a given data set. These values should be tuned so that the combination of hyper-parameters maximizes the model’s performance, minimizing a predefined loss function to produce better results with fewer errors(Probst et al., 2018; Weerts et al., 2020).

Since all four types of clouds will be unevenly distributed in the unseen test data, we use several evaluation metrics such as balanced accuracy(BAC), precision, recall and F1-score which are commonly used in classification tasks ( Silva et al., 2020; Nasir et al., 2022). A confusion matrix (Fig 6) of size n x n associated with a classifier shows the predicted and actual classification, where n is the number of different classes (Visa et al., 2011).

\section{RESULTS AND DISCUSSIONS}

\subsection{ Evolution case of a convective system} 

Figures 7, 8 and 9 depict the temporal evolution of a convective cloud  into stratiform precipitation over MDV. This rainfall event occurred on 26 June 2018 and was associated with a mesoscale convective system (MCS). An MCS comprises of mainly convective and stratiform regions and has a life cycle consisting of formation, maturity and dissipation (Houze 2004). A region of transitional clouds also exist in the MCS which forms when convective clouds begin to dissipate and merge into the stratiform anvil cloud (Smull and Houze 1987; Williams et al. 1995) . Here, initially, a fully developed convective cloud is seen (Fig. 7a) which starts to decay and transitions to mixed stratiform-convective system (Fig. 7b). At this stage, a local maximum in Z (indicative of bright band) starts to appear between 3 and 4 km height levels and become more pronounced in Fig. 7c. This suggests the gradual dissipation of convective features and shift towards stratiform cloud. The bright-band signature becomes well-marked in Fig. 7d signifying the transition to stratiform system is complete. Next, the stratiform system matures and eventually diminishes (Fig. 7e,f).

Corresponding vertical profiles of DV and W are shown in Fig. 8 and Fig. 9 respectively. During the transition phase, a gradient in DV emerges at bright band levels (3-4 km) (Fig. 8b,c) due to the increase in fall velocities of rain drops toward the end of melting process (Battan 1973). This gradient increases and peaks for mature stratiform system (Fig. 8d) when DV rises sharply from 2 at 3.5 km to 9 m/s at 2.5 km. In later stages, DV decreases with the dissipation of stratiform cloud (Fig. 8e,f). In case of W, enhanced values around 2-3 m/s are noted above the bright band during convective and transition phases indicating higher turbulence (Fig. 9a,b,c). Later, W decreases at these levels for stratiform system (Fig. 9d, e, f ). Williams et al. (1995) also noticed enhanced turbulence above the melting level (see their Fig. 6) in case of mixed clouds and reasoned that large W result from convective updrafts and downdrafts occurring at these height levels which is absent in stratiform system. An increase in W is also seen just below the bright band during the transition and stratiform stages which is associated with the increasing drop size as hydrometeors transform shape as they fall through the melting layer (Williams et al 1995).

\subsection{CloudSense Results}

We trained 7 ML models for this multi-classification task using the training dataset described in Section 2.2.3. Among these, 5 of them were hyper-parameter tuned models and 2 of them were not-tuned models. The ML algorithms used are Decision tree, Naïve Bayes, xgboost (Extreme Gradient Boosting), catboost (Categorical Boosting), KNN (K Nearest Neighbours), Random Forest and Light GBM (Light Gradient Boosting Machine). The test set consists of 400 clouds (stratiform-140, mixed-77, convective-99,  shallow -84) taken over the HACPL and MDV. The overall results obtained on the test set from the models are given in  Table 2.

It can be seen that hyper-parameter tuned models perform better than not-tuned models for this classification problem (Table 2). This is in agreement with the results of Ghada W et al., 2022. LightGBM model gives the best results out of the 5 tuned models with a BAC of 0.80 and F1-Score of 0.82. Therefore, this ML model is considered for CloudSense. Table 3 and Table 4 provide the classification report of CloudSense and the tuned hyper-parameter values for the LightGBM model.

Shallow clouds, as mentioned earlier, are correctly predicted with highest percentage and show the highest scores for precision, recall and F1-Score. Stratiform clouds show an F1-Score of 0.84 followed by convective clouds and mixed clouds with 0.78 and 0.66 respectively. This shows that the distribution of stratiform clouds is better understood by the model than convective and mixed clouds. To get the correctly classified and misclassified number of clouds by the model we analyse the confusion matrix (Fig 10).

Fig. 10 shows that 123 stratiform clouds (88 \%), 45 mixed clouds (58 \%), 80 convective clouds (81 \%), and 80 shallow clouds (95\%) have been accurately identified and classified by the model. Although the number of clouds correctly predicted by the model is considerable, there are instances of misclassifications. Specifically, 10 convective clouds have been classified as stratiform, and 12 stratiform clouds have been incorrectly classified as convective clouds. Various factors constrain the performance of the model such as limited sample size, data imbalance, relevance of features selected for training the ML model and other external factors. To understand the nature of misclassifications by the model, we test the model using a RHI scan conducted by the radar at 3:52 UTC on July 15, 2018 (Fig.11). 

Almost 75 \% of the precipitating cloud profiles have been correctly identified and classified. While shallow clouds have been predicted with 100 \% accuracy, there are some instances of misclassification with other cloud types. The misclassifications primarily involve stratiform and convective clouds being predicted as mixed stratiform-convective clouds, and vice versa. Notably, no convective cloud is predicted as stratiform or vice versa. Classifying a cloud type accurately is an important step in QPE. A stratiform cloud is considered to produce low intensity rain while a convective cloud is associated with heavy rain; rainfall from a mixed cloud is moderate in nature. Thus, these misclassifications directly impact the accuracy of QPE. However, the impact is relatively less if a stratiform/convective cloud is categorized as mixed and vice versa. On the other hand, misclassifying a convective cloud as stratiform can have severe consequences, leading to forecasts with significantly underestimated rainfall. It may be noted that misclassification of this nature has not been observed here; however, it cannot be completely ruled out for other cases. In summary, the model produced cloud types have instances of misclassifications but with relatively less impact and therefore, the performance of the model may be considered satisfactory.

\subsection{ CloudSense versus radar algorithm output}

In this section, we compare the statistics of CloudSense results with those of a conventional radar algorithm. Williams et al. (1995) have utilized vertical profiles of Z, DV and W collected using a vertically pointing 915 MHz wind profiler radar to classify clouds into the four types.

They used 3 criteria to distinguish the 4 cloud types (see their Fig 4): melting layer signature, turbulence above the melting level and hydrometeors above the melting level. A set of thresholds were specified by them for maximum Doppler velocity gradient (DVG) between the height levels of 3.5 and 5 km and maximum spectral width (MSW) above 7 km. 

In this work, we have adapted their methodology and adjusted their algorithm to account for the differences in the observing instrument and the region of study.  For example, their 915 MHz wind profiler radar is sensitive to both turbulence in clear air and precipitating hydrometeors while the X-band radar used here is sensitive to precipitating particles only and provides data at comparatively higher resolution. Additionally, there are differences in melting layer altitudes at both sites. A  comparison of the CFADs of Z, DV and W constructed over MDV (Fig. 12) with those of Williams et al. (1995) (their Fig. 3) highlight these differences. Considering these distinctions, we set thresholds of DVG between height levels of 2.8-4 km and that of MSW between 4-7 km.

A total of 200 VPRs collected from the MDV site with 50 profiles each for all four cloud types are taken for testing the algorithm. We tested the sensitivity of classification results for a set of threshold values for DVG and MSW (refer Table 1 in supplementary document) and found that a DVG $>$ 2.5 m s\textsuperscript{-1} km\textsuperscript{-1} and MSW $>$ 2 m/s yielded the best results. Corresponding to this threshold, the evaluation metrics are presented in Table 5 and 6 for radar algorithm and CloudSense model, respectively. 

The radar algorithm exhibited a BAC of 0.69 and F1-Score of 0.68, whereas CloudSense achieved a BAC and F1-Score of 0.77, indicating that the CloudSense model outperforms the threshold-based radar algorithm. Note that as shallow clouds are classified using the same criteria for both the model and radar algorithm, the classification results are similar for both.

\section{SUMMARY AND CONCLUSIONS}

This paper presents an initial attempt to develop an accurate cloud classification model from radar data with an ML-based approach which is completely lacking over the Indian region. We proposed a novel model called CloudSense which uses machine learning (ML) for identifying precipitating cloud types over the WGs. The clouds are classified into stratiform, mixed stratiform-convective, convective and shallow. The model used vertical reflectivity profiles (VPRs) collected from an X-band DWR located at Mandhardev (18.04°N, 73.86°E and $\sim$ 1.3km MSL) in the WGs. The CloudSense model classifies shallow clouds using flexible threshold height of the melting layer whereas other 3 cloud types are classified using ML algorithms. A typical case showing the evolution of a convective system to stratiform precipitation was discussed and microphysical properties of the clouds were studied to select the suitable features to train the ML algorithm. 7 ML algorithms were trained and tested using the VPRs. Among all ML models, Light GBM, a tree-based learning algorithm, showed the best test results with a BAC of 0.8 and F1-Score of 0.82. The performance of CloudSense was compared against a conventional threshold-based radar algorithm described in by Williams et al 1995. The radar algorithm achieved a BAC of 0.69 and F1-Score of 0.68 while the model exhibited a BAC and F1-Score of 0.77 for 200 samples tested, thus showing that CloudSense can provide more accurate detection and classification of cloud types.

Future plan is to rigorously train this model using more samples and develop a more robust model for improved classification of clouds over the WG. Similar studies can be carried using other radars over other regions to develop accurate cloud classification models which would be useful to improve the radar based quantitative estimates of precipitation.\\

 \textbf{Acknowledgements}: The Indian Institute of Tropical Meteorology (IITM), Pune, is an autonomous organization, fully funded by the Ministry of Earth Sciences (MoES), Government of India. The authors extend their gratitude to the Director, IITM for providing unwavering support and encouragement throughout this research. Author Mehzooz Nizar would love to thank Meenu R. Nair, Adithiy Raghunathan, his professors at Cochin University of Science and Technology, the Open Radar Science community and all his workmates at IITM for their invaluable support and help. The authors gratefully acknowledge the dedicated efforts of the technical staff at HACPL and the radar site for their diligent work in observations and maintenance.\newline\newline

\textbf{Funding}: This research did not receive any specific grant from funding agencies in the public, commercial, or not-for-profit sectors.\newline

\textbf{REFERENCES}\newline

Anagnostou, E.N., 2004. A convective/stratiform precipitation classification algorithm for volume scanning weather radar observations. Meteorol. Appl. 11, 291–300. 

Arulraj, M., Barros, A.P., 2019. Improving quantitative precipitation estimates in mountainous regions by modelling low-level seeder-feeder interactions constrained by Global Precipitation Measurement Dual-frequency Precipitation Radar measurements. Remote Sens. Environ. 231, 111213. 

Auipong, N., Trivej, P., 2018. Study of Z-R relationship among different topographies in Northern Thailand. J. Phys. Conf. Ser. 1144, 012098. 

Austin, P.M., Houze, R.A., 1972. Analysis of the Structure of Precipitation Patterns in New England. J. Appl. Meteorol. 11, 926–935. 

Barua, S., Islam, M.M., Murase, K., 2011. A Novel Synthetic Minority Oversampling Technique for Imbalanced Data Set Learning BT  - Neural Information Processing, in: Lu, B.-L., Zhang, L., Kwok, J. (Eds.), . Springer Berlin Heidelberg, Berlin, Heidelberg, pp. 735–744.

Battan, L.J., 1973. Radar observation of the atmosphere. L. J. Battan (The University of Chicago Press) 1973. PP X, 324; 125 figures, 21 tables. £7·15. Q. J. R. Meteorol. Soc. 99, 793–793. 

Biggerstaff, M.I., Listemaa, S.A., 2000. An Improved Scheme for Convective/Stratiform Echo Classification Using Radar Reflectivity. J. Appl. Meteorol. 39, 2129–2150. 

Biswas, S., Chandrasekar, V., 2018. Cross-Validation of Observations between the GPM Dual-Frequency Precipitation Radar and Ground Based Dual-Polarization Radars. Remote Sens. 10, 1773.

Chawla, N. V, Bowyer, K.W., Hall, L.O., Kegelmeyer, W.P., 2002. SMOTE: Synthetic Minority Over-sampling Technique. J. Artif. Intell. Res. 16, 321–357.  

Churchill, D.D., Houze, R.A., 1984. Development and Structure of Winter Monsoon Cloud Clusters On 10 December 1978. J. Atmos. Sci. 41, 933–960. 

Das, S.K., Konwar, M., Chakravarty, K., Deshpande, S.M., 2017. Raindrop size distribution of different cloud types over the Western Ghats using simultaneous measurements from Micro-Rain Radar and disdrometer. Atmos. Res. 186, 72–82. 

DeMott, C. A., R. Cifelli, and S. A. Rutledge, 1995: An improved method for partitioning radar data into convective and stratiform components. Preprints, 27th Conf. on Radar Meteorology, Vail, CO, Amer. Meteor. Soc., 233–236.

Deshpande, S.M., Dhangar, N., Das, S.K., Kalapureddy, M.C.R., Chakravarty, K., Sonbawne, S., Konwar, M., 2015. Mesoscale kinematics derived from X‐band Doppler radar observations of convective versus stratiform precipitation and comparison with GPS radiosonde profiles. J. Geophys. Res. Atmos. 120, 11,511-536,551. 

Devisetty, H.K., Jha, A.K., Das, S.K., Deshpande, S.M., Krishna, U.V.M., Kalekar, P.M., Pandithurai, G., 2019. A case study on bright band transition from very light to heavy rain using simultaneous observations of collocated X- and Ka-band radars. J. Earth Syst. Sci. 128, 136.

Elreedy, D., Atiya, A.F., 2019. A Comprehensive Analysis of Synthetic Minority Oversampling Technique (SMOTE) for handling class imbalance. Inf. Sci. (Ny). 505, 32–64. 

Fabry, F., 2015. Radar Meteorology. Cambridge University Press, Cambridge.  

Ghada, W., Casellas, E., Herbinger, J., Garcia-Benadí, A., Bothmann, L., Estrella, N., Bech, J., Menzel, A., 2022. Stratiform and Convective Rain Classification Using Machine Learning Models and Micro Rain Radar. Remote Sens. 14, 4563. 

Gourley, J.J., Vieux, B.E., 2005. A Method for Evaluating the Accuracy of Quantitative Precipitation Estimates from a Hydrologic Modeling Perspective. J. Hydrometeorol. 6, 115–133. 

Houze, Jr., R.A., 1982. Cloud Clusters and Large-Scale Vertical Motions in the Tropics. J. Meteorol. Soc. Japan. Ser. II 60, 396–410. 

Houze, R.A., 1997. Stratiform Precipitation in Regions of Convection: A Meteorological Paradox? Bull. Am. Meteorol. Soc. 78, 2179–2196. 

Houze, R.A., 1973. A Climatological Study of Vertical Transports by Cumulus-Scale Convection. J. Atmos. Sci. 30, 1112–1123. 

Houze, R.A., 2004. Mesoscale convective systems. Rev. Geophys. 42.  

Houze, R.A., 2014. Cloud Dynamics - Second Edition, International Geophysics.

Inoue, T., 1987. A cloud type classification with NOAA 7 split‐window measurements. J. Geophys. Res. Atmos. 92, 3991–4000. 

Jha, A.K., Kalapureddy, M.C.R., Devisetty, H.K., Deshpande, S.M., Pandithurai, G., 2019. A case study on large-scale dynamical influence on bright band using cloud radar during the Indian summer monsoon. Meteorol. Atmos. Phys. 131, 505–515. 

Kalapureddy, M.C.R., Patra, S., Dhavale, V., Nair, M.R., 2023. CloudSat inferred contrasting monsoon intra-seasonal variation in the cloud vertical structure over Indian regions. Clim. Dyn. 61, 1567–1589. 

Konwar, M., Das, S.K., Deshpande, S.M., Chakravarty, K., Goswami, B.N., 2014. Microphysics of clouds and rain over the Western Ghat. J. Geophys. Res. Atmos. 119, 6140–6159. 

Konwar, M., Maheskumar, R.S., Kulkarni, J.R., Freud, E., Goswami, B.N., Rosenfeld, D., 2012. Aerosol control on depth of warm rain in convective clouds. J. Geophys. Res. Atmos. 117. 

Kühnlein, M., Appelhans, T., Thies, B., Nauss, T., 2014. Improving the accuracy of rainfall rates from optical satellite sensors with machine learning — A random forests-based approach applied to MSG SEVIRI. Remote Sens. Environ. 141, 129–143. 

Kumar Das, S., Deshpande, S.M., Shankar Das, S., Konwar, M., Chakravarty, K., Kalapureddy, M.C.R., 2015. Temporal and structural evolution of a tropical monsoon cloud system: A case study using X-band radar observations. J. Atmos. Solar-Terrestrial Phys. 133, 157–168. 

Lavanya, S., Kirankumar, N.V.P., 2021. Classification of tropical coastal precipitating cloud systems using disdrometer observations over Thumba, India. Atmos. Res. 253, 105477. 

Lazri, M., Ameur, S., 2018. Combination of support vector machine, artificial neural network and random forest for improving the classification of convective and stratiform rain using spectral features of SEVIRI data. Atmos. Res. 203, 118–129. 

Loh, J. Le, Lee, D.-I., Kang, M.-Y., You, C.-H., 2020. Classification of Rainfall Types Using Parsivel Disdrometer and S-Band Polarimetric Radar in Central Korea. Remote Sens. 12, 642. 

Maheskumar, R.S., Narkhedkar, S.G., Morwal, S.B., Padmakumari, B., Kothawale, D.R., Joshi, R.R., Deshpande, C.G., Bhalwankar, R. V, Kulkarni, J.R., 2014. Mechanism of high rainfall over the Indian west coast region during the monsoon season. Clim. Dyn. 43, 1513–1529. 

Marshall, J.S., Langille, R.C., Palmer, W.M.K., 1947. MEASUREMENT OF RAINFALL BY RADAR. J. Meteorol. 4, 186–192.  

McKee, J.L., Binns, A.D., 2016. A review of gauge–radar merging methods for quantitative precipitation estimation in hydrology. Can. Water Resour. J. / Rev. Can. des ressources hydriques 41, 186–203. 

Nandargi, S., Mulye, S.S., 2012. Relationships between Rainy Days, Mean Daily Intensity, and Seasonal Rainfall over the Koyna Catchment during 1961–2005. Sci. World J. 2012, 1–10.  

Nasir, N., Kansal, A., Alshaltone, O., Barneih, F., Sameer, M., Shanableh, A., Al-Shamma’a, A., 2022. Water quality classification using machine learning algorithms. J. Water Process Eng. 48, 102920. 

Özöğür-Akyüz, S., Ünay, D., Smola, A., 2011. Guest editorial: model selection and optimization in machine learning. Mach. Learn. 85, 1–2. 

Penide, G., Protat, A., Kumar, V. V, May, P.T., 2013. Comparison of Two Convective/Stratiform Precipitation Classification Techniques: Radar Reflectivity Texture versus Drop Size Distribution–Based Approach. J. Atmos. Ocean. Technol. 30, 2788–2797. 

Probst, P., Boulesteix, A., Bischl, B., 2018. Tunability: Importance of Hyperparameters of Machine Learning Algorithms. J. Mach. Learn. Res. 20, 53:1-53:32.

Ran, Y., Wang, H., Tian, L., Wu, J., Li, X., 2021. Precipitation cloud identification based on faster-RCNN for Doppler weather radar. EURASIP J. Wirel. Commun. Netw. 2021, 19. 

Rao, T.N., Rao, D.N., Mohan, K., Raghavan, S., 2001. Classification of tropical precipitating systems and associated Z ‐ R relationships. J. Geophys. Res. Atmos. 106, 17699–17711.  

Sassen, K., Wang, Z., 2008. Classifying clouds around the globe with the CloudSat radar: 1‐year of results. Geophys. Res. Lett. 35. 

Schumacher, C., Houze, R.A., 2003. The TRMM Precipitation Radar’s View of Shallow, Isolated Rain. J. Appl. Meteorol. 42, 1519–1524. 

Silva, A.A., Tavares, M.W., Carrasquilla, A., Misságia, R., Ceia, M., 2020. Petrofacies classification using machine learning algorithms. GEOPHYSICS 85, WA101–WA113. 

Smull, B.F., Houze, R.A., 1987. Dual-Doppler Radar Analysis of a Midlatitude Squall Line with a Trailing Region of Stratiform Rain. J. Atmos. Sci. 44, 2128–2149. 

So, D., Shin, D.-B., 2018. Classification of precipitating clouds using satellite infrared observations and its implications for rainfall estimation. Q. J. R. Meteorol. Soc. 144, 133–144. 

Steiner, M., Houze, R.A., 1997. Sensitivity of the Estimated Monthly Convective Rain Fraction to the Choice of Z – R Relation. J. Appl. Meteorol. 36, 452–462. 

Steiner, M., Houze, R.A., Yuter, S.E., 1995. Climatological Characterization of Three-Dimensional Storm Structure from Operational Radar and Rain Gauge Data. J. Appl. Meteorol. 34, 1978–2007. 

Subrahmanyam, K. V, Kumar, K.K., 2013. CloudSat observations of cloud-type distribution over the Indian summer monsoon region. Ann. Geophys. 31, 1155–1162.  

Testud, J., Oury, S., Black, R.A., Amayenc, P., Dou, X., 2001. The Concept of “Normalized” Distribution to Describe Raindrop Spectra: A Tool for Cloud Physics and Cloud Remote Sensing. J. Appl. Meteorol. 40, 1118–1140. 

Thompson, E.J., Rutledge, S.A., Dolan, B., Thurai, M., 2015. Drop Size Distributions and Radar Observations of Convective and Stratiform Rain over the Equatorial Indian and West Pacific Oceans. J. Atmos. Sci. 72, 4091–4125. 

Tokay, A., Short, D.A., 1996. Evidence from Tropical Raindrop Spectra of the Origin of Rain from Stratiform versus Convective Clouds. J. Appl. Meteorol. 35, 355–371. 

Visa, S., Ramsay, B., Ralescu, A., Knaap, E., 2011. Confusion Matrix-based Feature Selection., CEUR Workshop Proceedings.

Waldvogel, A., 1974. The N 0 Jump of Raindrop Spectra. J. Atmos. Sci. 31, 1067–1078. 

Wang, H., Shao, N., Ran, Y., 2019. Identification of Precipitation-Clouds Based on the Dual-Polarization Doppler Weather Radar Echoes Using Deep–Learning Method. IEEE Access 7, 12822–12831. 

Wang, Y., Tang, L., Chang, P.-L., Tang, Y.-S., 2021. Separation of convective and stratiform precipitation using polarimetric radar data with a support vector machine method. Atmos. Meas. Tech. 14, 185–197. 

Weerts, H.J.P., Mueller, A., Vanschoren, J., 2020. Importance of Tuning Hyperparameters of Machine Learning Algorithms. ArXiv abs/2007.0.

Williams, C.R., Ecklund, W.L., Gage, K.S., 1995. Classification of Precipitating Clouds in the Tropics Using 915-MHz Wind Profilers. J. Atmos. Ocean. Technol. 12, 996–1012. 

Yang, Z., Liu, P., Yang, Y., 2019. Convective/Stratiform Precipitation Classification Using Ground-Based Doppler Radar Data Based on the K-Nearest Neighbor Algorithm. Remote Sens. 11, 2277. 

Yan-Jiao, X., Li-Ping, L.I.U., 2007. Identification of Stratiform and Convective Cloud Using 3D Radar Reflectivity Data. Chinese J. Atmos. Sci. 31, 645–654. 

Yuter, S.E., Houze, R.A., 1995. Three-Dimensional Kinematic and Microphysical Evolution of Florida Cumulonimbus. Part II: Frequency Distributions of Vertical Velocity, Reflectivity, and Differential Reflectivity. Mon. Weather Rev. 123, 1941–1963. 

Zhong, L., Yang, R., Wen, Y., Chen, L., Gou, Y., Li, R., Zhou, Q., Hong, Y., 2017. Cross-evaluation of reflectivity from the space-borne precipitation radar and multi-type ground-based weather radar network in China. Atmos. Res. 196, 200–210. 

Zhu, T., Lin, Y., Liu, Y., 2017. Synthetic minority oversampling technique for multiclass imbalance problems. Pattern Recognit. 72, 327–340. 

Zuo, Y., Hu, Z., Yuan, S., Zheng, J., Yin, X., Li, B., 2022. Identification of Convective and Stratiform Clouds Based on the Improved DBSCAN Clustering Algorithm. Adv. Atmos. Sci. 39, 2203–2212.


\begin{table}[hbt!]
\centering
\caption{\textbf{Technical specifications of X-band Doppler radar}}
\begin{tabular}{| l | l |}
\hline
\textbf{Parameter} & \textbf{Value} \\
\hline
\textbf{Frequency(GHz)} & \textbf{9.535} \\
\hline
\textbf{Wavelength(cm)} & \textbf{3.14} \\
\hline
\textbf{Transmitter} & \textbf{Magnetron} \\
\hline
\textbf{Peak power(kW)} & \textbf{200} \\
\hline
\textbf{Pulse widths} & \textbf{0.8-2 µs} \\
\hline
\textbf{Beam width} &  \\
\hline
\textbf{Antenna gain(dB)} & \textbf{44.3} \\
\hline
\textbf{Antenna diameter(m)} & \textbf{2.4} \\
\hline
\textbf{Cross-pol Isolation(dB)} & \textbf{-30} \\
\hline
\textbf{Minimum detectable signal} & \textbf{-25 dBZ at 20 km} \\
\hline

\end{tabular}

\end{table}

\begin{table}[hbt!]
\centering
\caption{Overall results of tuned and not tuned ML models. The bold and underlined portion highlights the best results}
\begin{tabular}{| l | l | l | l | l | l |}
\hline
 & Model & BAC & Precision & Recall & F1-Score \\
\hline
\multirow{2}{*}{Not Tuned Models} & Decision tree & 0.67 & 0.68 & 0.68 & 0.68 \\
 & Naïve Bayes & 0.54 & 0.69 & 0.43 & 0.36 \\
\multirow{5}{*}{Tuned Models} & xgboost & 0.76 & 0.78 & 0.79 & 0.78 \\
\hline
 & catboost & 0.76 & 0.78 & 0.78 & 0.77 \\
 & KNN & 0.69 & 0.71 & 0.71 & 0.71 \\
 & \textbf{Light GBM} & \textbf{0.80} & \textbf{0.82} & \textbf{0.82} & \textbf{0.82} \\
 & Random Forest & 0.77 & 0.79 & 0.79 & 0.79 \\ \hline

\end{tabular}

\end{table}

\begin{table}[hbt!]
\centering
\caption{\textbf{ Classification report of CloudSense }}
\begin{tabular}{| l | l | l | l | l |}
\hline
 & Precision & Recall & F1-Score & Number of samples \\
\hline
Stratiform & 0.80 & 0.88 & 0.84 & 140 \\
\hline
Mixed Stratiform-Convective & 0.75 & 0.58 & 0.66 & 77 \\
\hline
Convective & 0.75 & 0.81 & 0.78 & 99 \\
\hline
Shallow & 1 & 0.95 & 0.98 & 84 \\
\hline

\end{tabular}

\end{table}

\begin{table}[hbt!]
\centering
\caption{\textbf{ Hyper-parameters and their tuned values used in the LightGBM model. The values are finalised after running 1000 trials and selecting the trial with the highest F1-Score}}
\begin{tabular}{| l | l | l | l |}
\hline
\textbf{Hyper Parameters} & \textbf{Lower limit} & \textbf{Upper limit} & \textbf{Tuned Values} \\
\hline
\textbf{boosting\_type} & \textbf{-} & \textbf{-} & \textbf{gbdt} \\
\hline
\textbf{class\_weight} & \textbf{1} & \textbf{5} & \textbf{\{0: 5, 1: 5, 2: 4\}} \\
\hline
\textbf{objective} & \textbf{-} & \textbf{-} & \textbf{multiclass} \\
\hline
\textbf{metric} & \textbf{-} & \textbf{-} & \textbf{multi\_logloss} \\
\hline
\textbf{random\_state} & \textbf{-} & \textbf{-} & \textbf{42} \\
\hline
\textbf{lambda\_l1} & \textbf{1e-8} & \textbf{10.0} & \textbf{3.54e-05} \\
\hline
\textbf{lambda\_l2} & \textbf{1e-8} & \textbf{10.0} & \textbf{0.0006} \\
\hline
\textbf{feature\_fraction} & \textbf{0.4} & \textbf{1} & \textbf{0.72} \\
\hline
\textbf{bagging\_fraction} & \textbf{0.4} & \textbf{1} & \textbf{0.98} \\
\hline
\textbf{bagging\_freq} & \textbf{1} & \textbf{7} & \textbf{6} \\
\hline
\textbf{min\_child\_samples} & \textbf{5} & \textbf{100} & \textbf{16} \\
\hline

\end{tabular}

\end{table}

\begin{table}[hbt!]
\centering
\caption{\textbf{Classification report of the radar algorithm for 200 cloud samples taken over the MDV site}}
\begin{tabular}{| l | l | l | l | l |}
\hline
 & Precision & Recall & F1-Score & Number of samples \\
\hline
Stratiform & 0.65 & 0.84 & 0.73 & 50 \\
\hline
Mixed Stratiform-Convective & 0.44 & 0.34 & 0.38 & 50 \\
\hline
Convective & 0.64 & 0.64 & 0.64 & 50 \\
\hline
Shallow & 1 & 0.92 & 0.96 & 50 \\
\hline

\end{tabular}

\end{table}

\begin{table}[hbt!]
\centering
\caption{\textbf{Classification report of CloudSense for 200 cloud samples taken over the MDV site}}
\begin{tabular}{| l | l | l | l | l |}
\hline
 & Precision & Recall & F1-Score & Number of samples \\
\hline
Stratiform & 0.68 & 0.80 & 0.73 & 50 \\
\hline
Mixed Stratiform-Convective & 0.81 & 0.50 & 0.62 & 50 \\
\hline
Convective & 0.67 & 0.86 & 0.75 & 50 \\
\hline
Shallow & 1 & 0.92 & 0.96 & 50 \\
\hline

\end{tabular}

\end{table}


\begin{figure}[hbt!]
    \centering
    \includegraphics[width=0.75\linewidth]{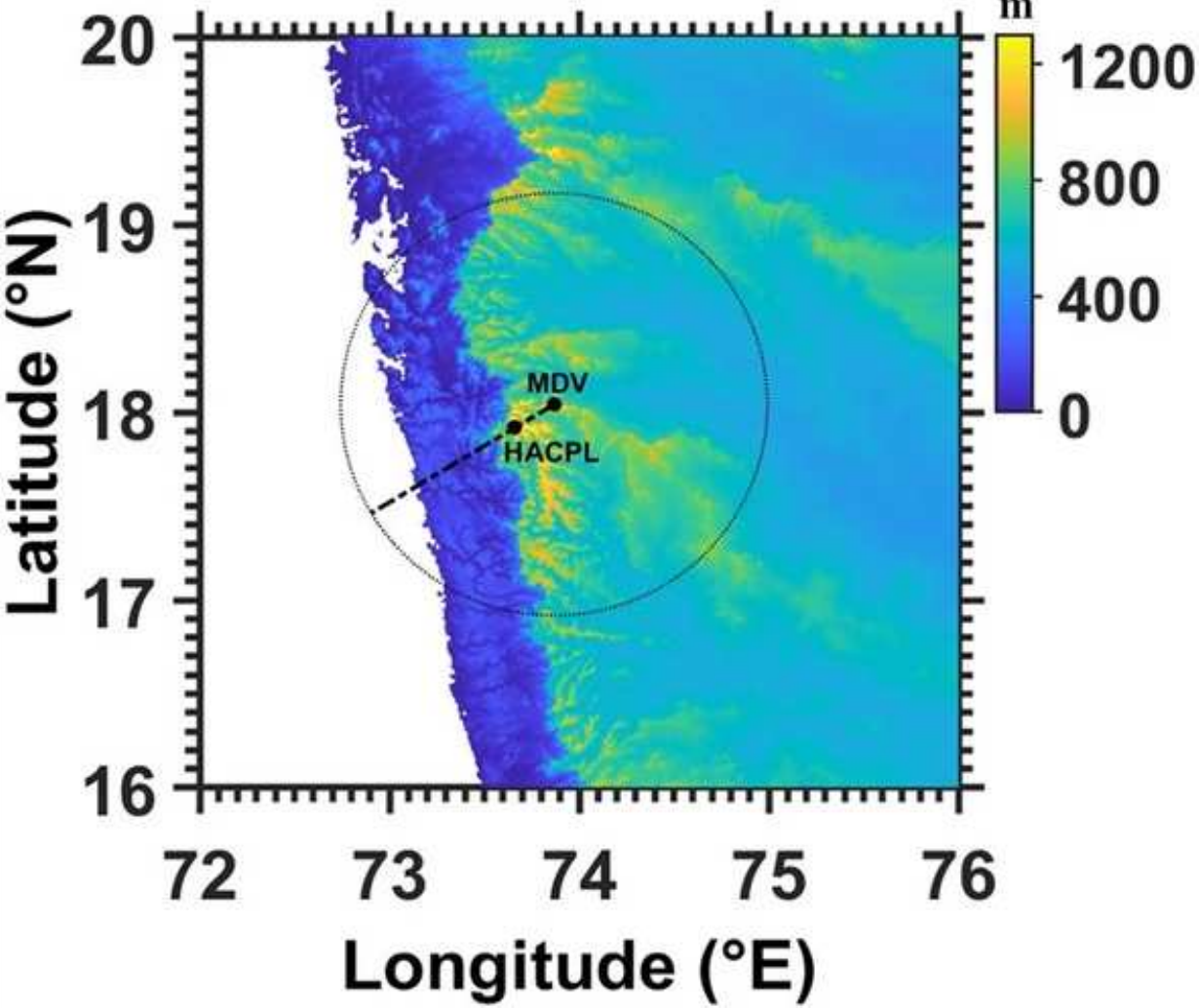}
    \caption{\textbf{ Topography map of the Western Ghats. Black dots show the locations of radar site (MDV) and HACPL. Dotted circle represents the X-band radar's maximum range (125 km). RHI scan is taken along the dash-dotted line}}

\end{figure}

\begin{figure}[hbt!]
    \centering
    \includegraphics[width=1\linewidth]{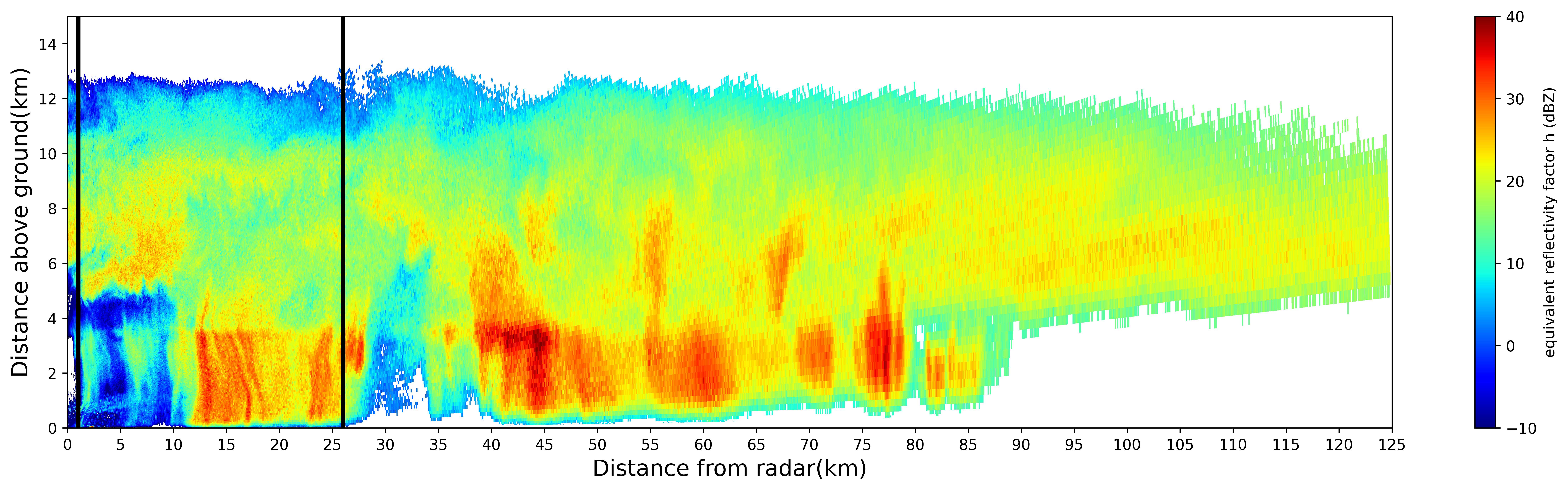}
    \label{FIG 2 a}
    
\end{figure}
\begin{figure}[hbt!]
    \centering
    \includegraphics[width=1\linewidth]{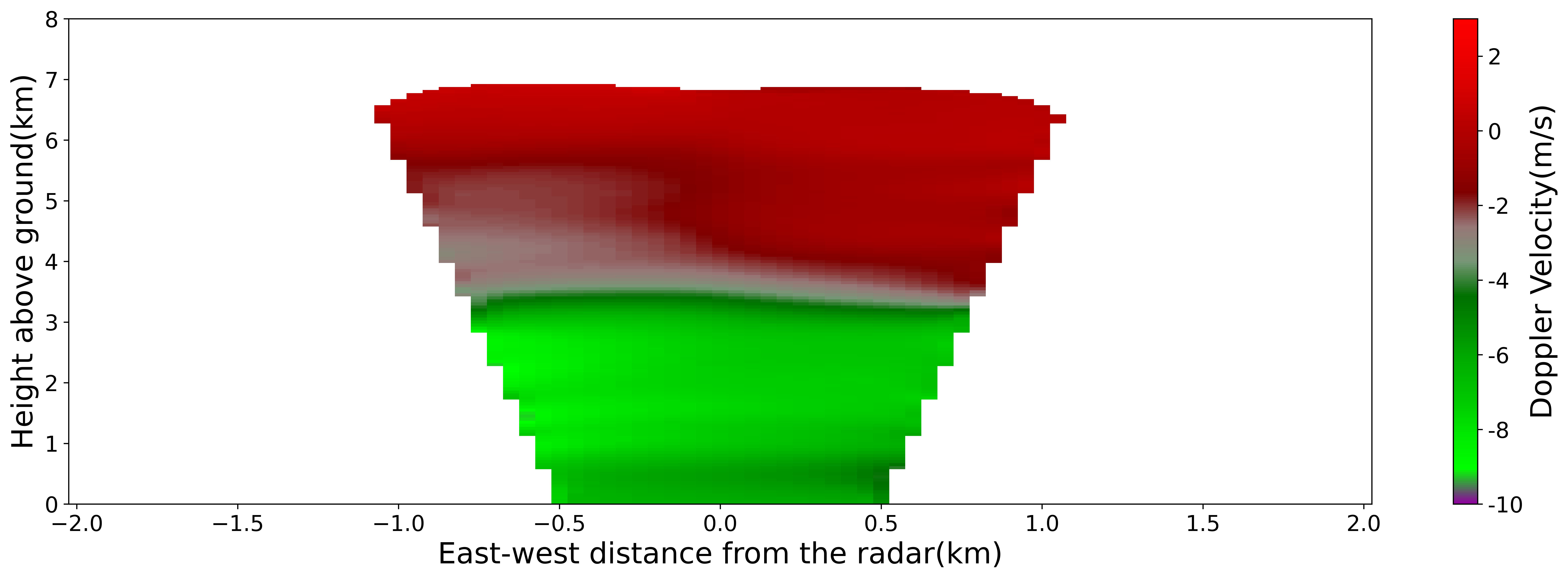}
    \caption{\textbf{ a) Typical RHI scan observed on X July, 2018. Vertical black line is drawn at 26 km to show VPRs considered over HACPL. b) Vertical profile of Doppler velocity from a PPI scan at 85º elevation}
}
    \label{FIG 2 b}
\end{figure}
\begin{figure}[hbt!]
    \centering
    \includegraphics[width=1\linewidth]{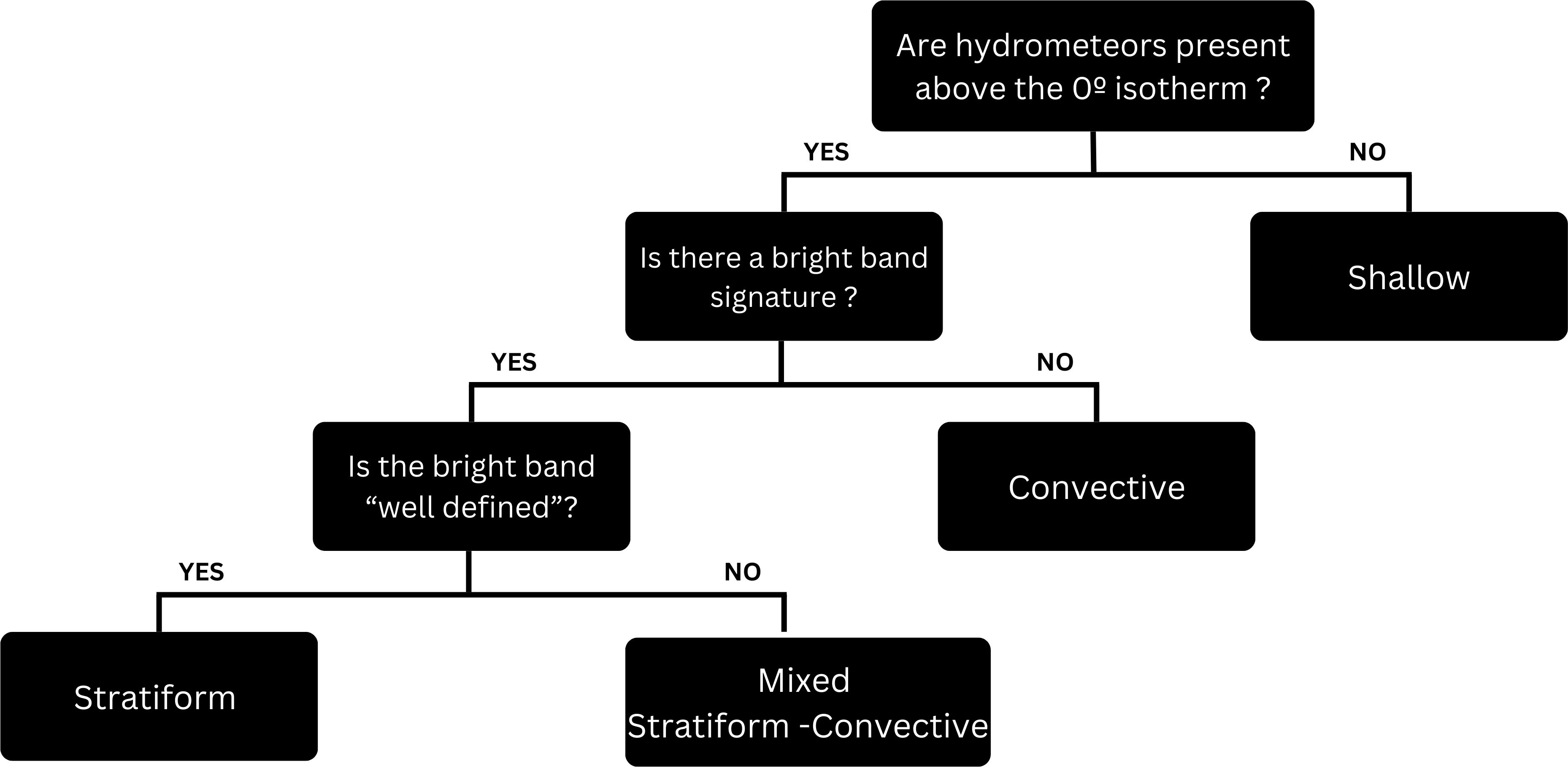}
    \caption{\textbf{ Logical flow diagram for classifying clouds}
}
    \label{FIG 3}
\end{figure}
\begin{figure}[hbt!]
    \centering
    \includegraphics[width=1\linewidth]{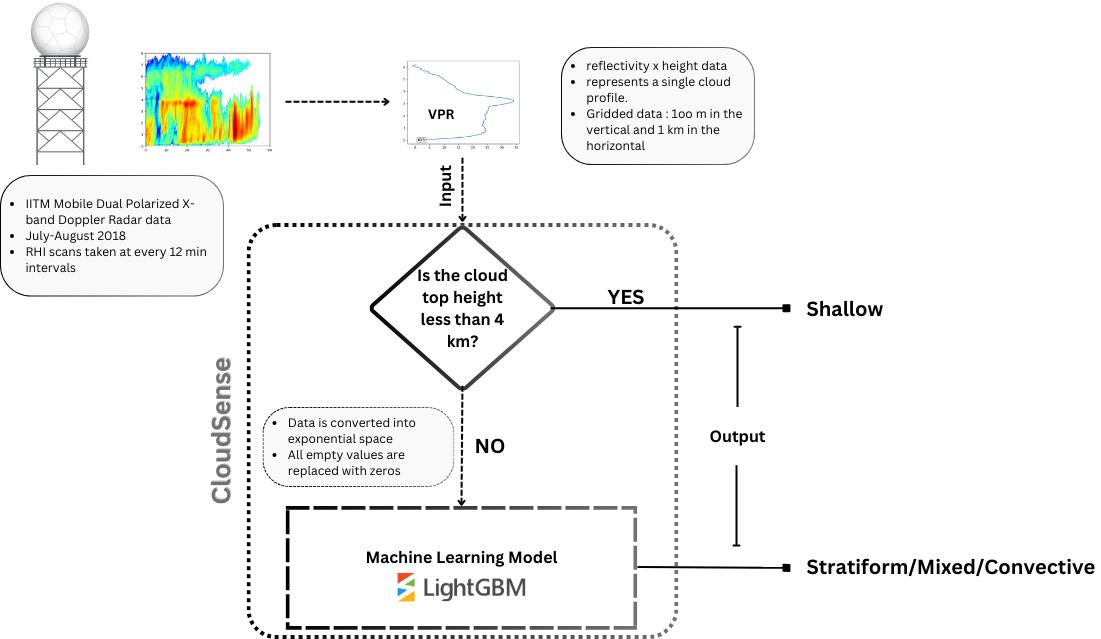}
    \caption{\textbf{ Architecture of CloudSense}
}
    \label{FIG 4}
\end{figure}
\begin{figure}[hbt!]
    \centering
    \includegraphics[width=1\linewidth]{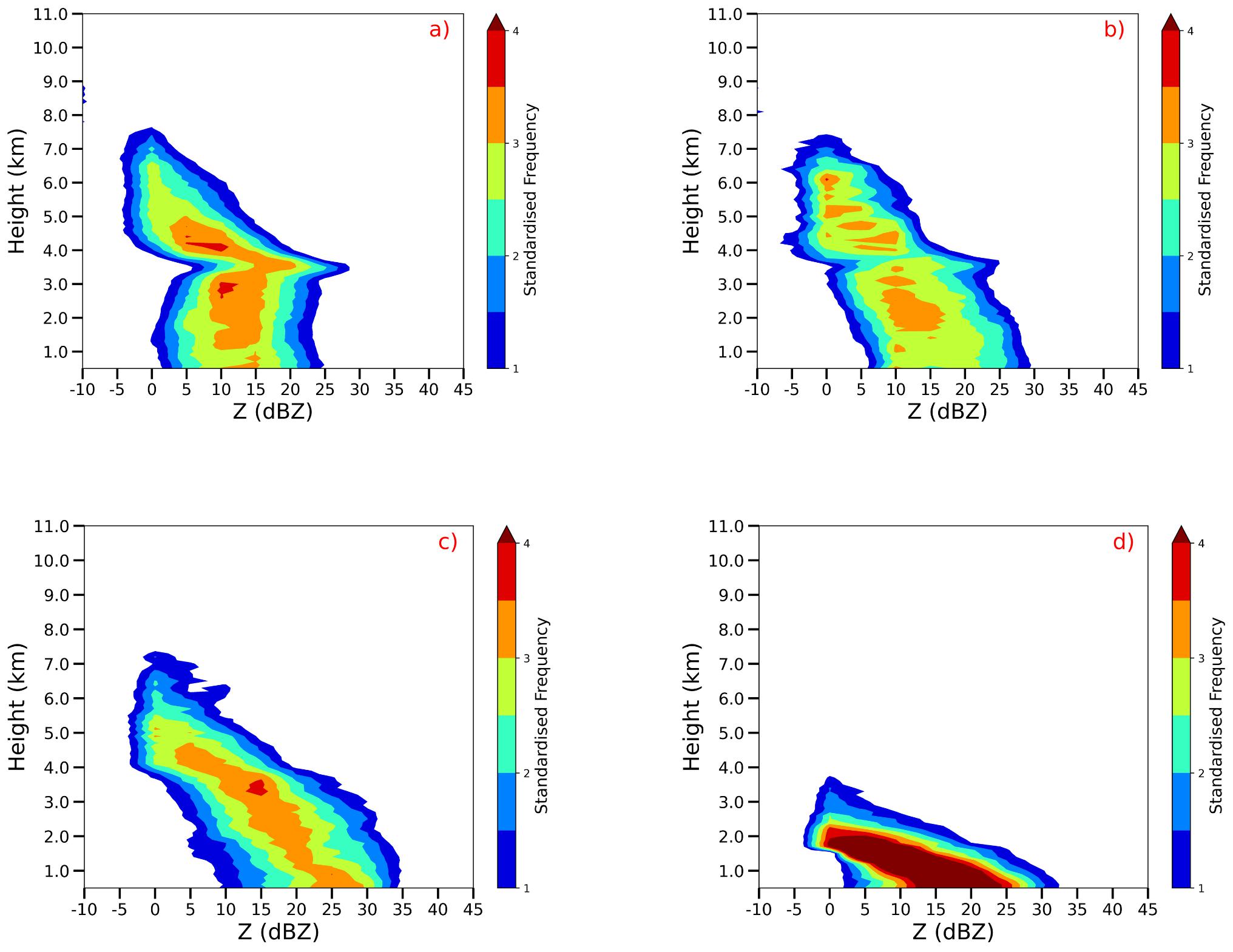}
    \caption{\textbf{ CFADs of Z for a) Stratiform b) Mixed Stratiform-Convective c) Convective and d)Shallow clouds}
}
    \label{FIG 5}
\end{figure}
\begin{figure}
    \centering
    \includegraphics[width=0.75\linewidth]{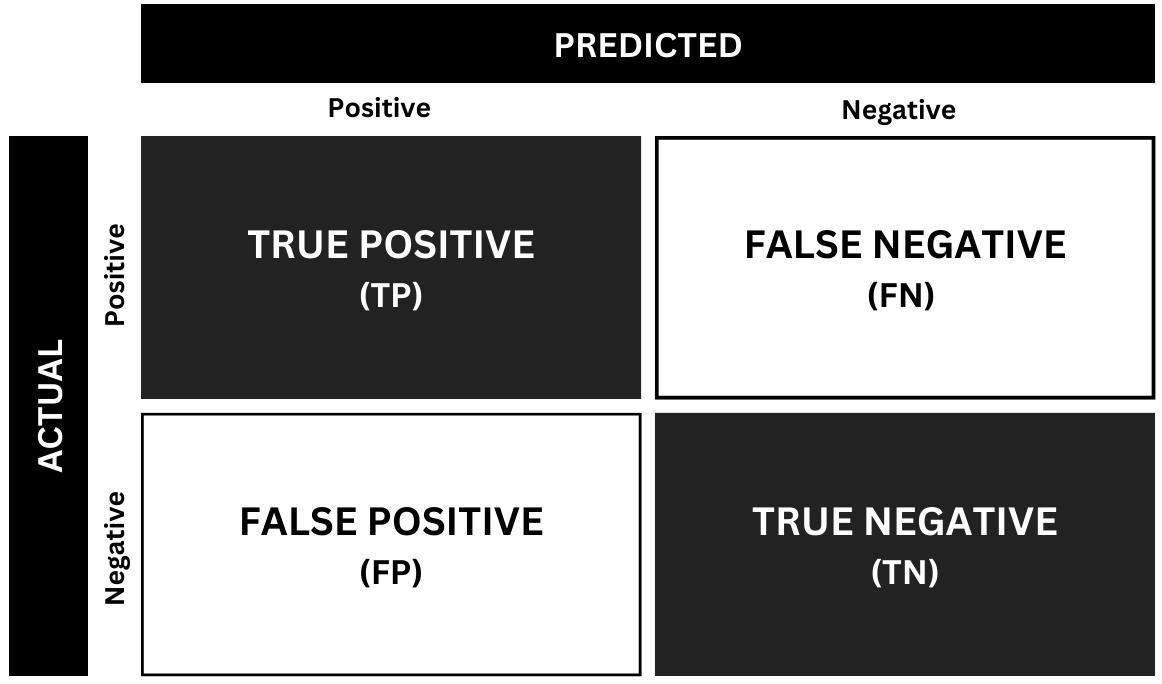}
    
    \caption{\textbf{A 2X2 Confusion Matrix} }
    \label{FIG 6}
\end{figure}
\begin{figure}[hbt!]
    \centering
 
    \renewcommand{\thefigure}{7}
    \includegraphics[width=1\linewidth]{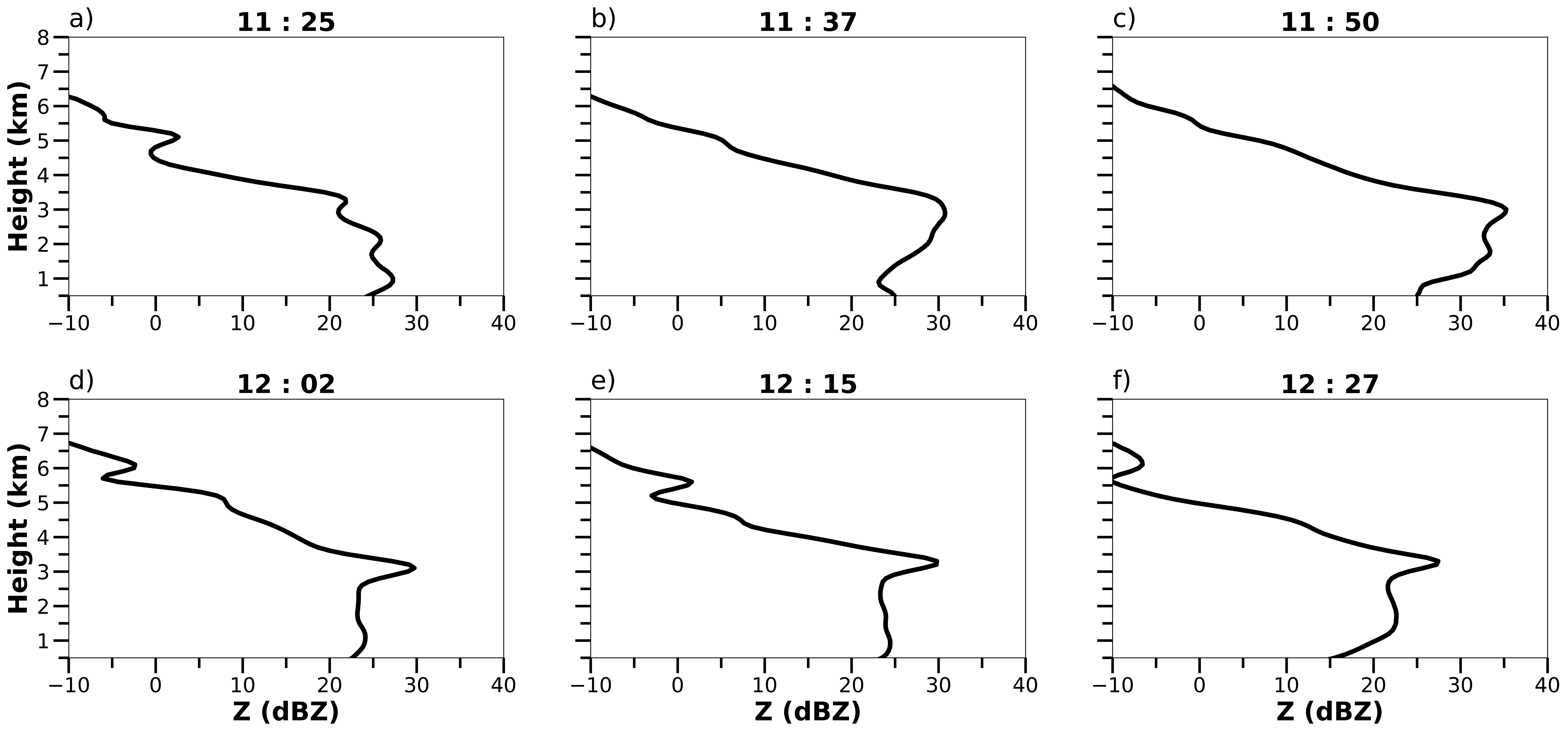}
    \label{FIG 7}
    
\end{figure}
\begin{figure}[hbt!]
    \centering
      \renewcommand{\thefigure}{8}
    \includegraphics[width=1\linewidth]{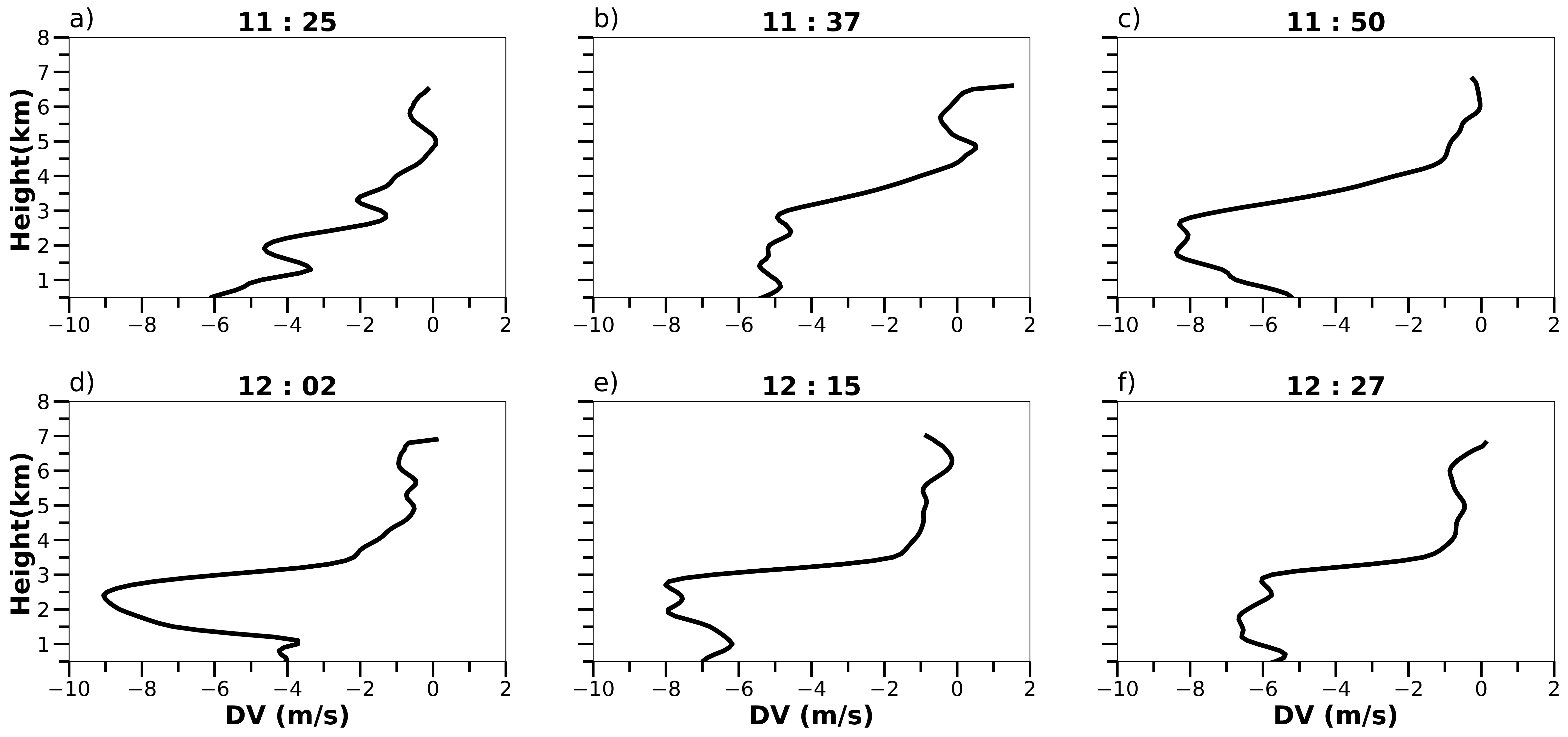}
    \label{FIG 8}

\end{figure}
\begin{figure}[hbt!]
    \centering
    \renewcommand{\thefigure}{9}
    \includegraphics[width=1\linewidth]{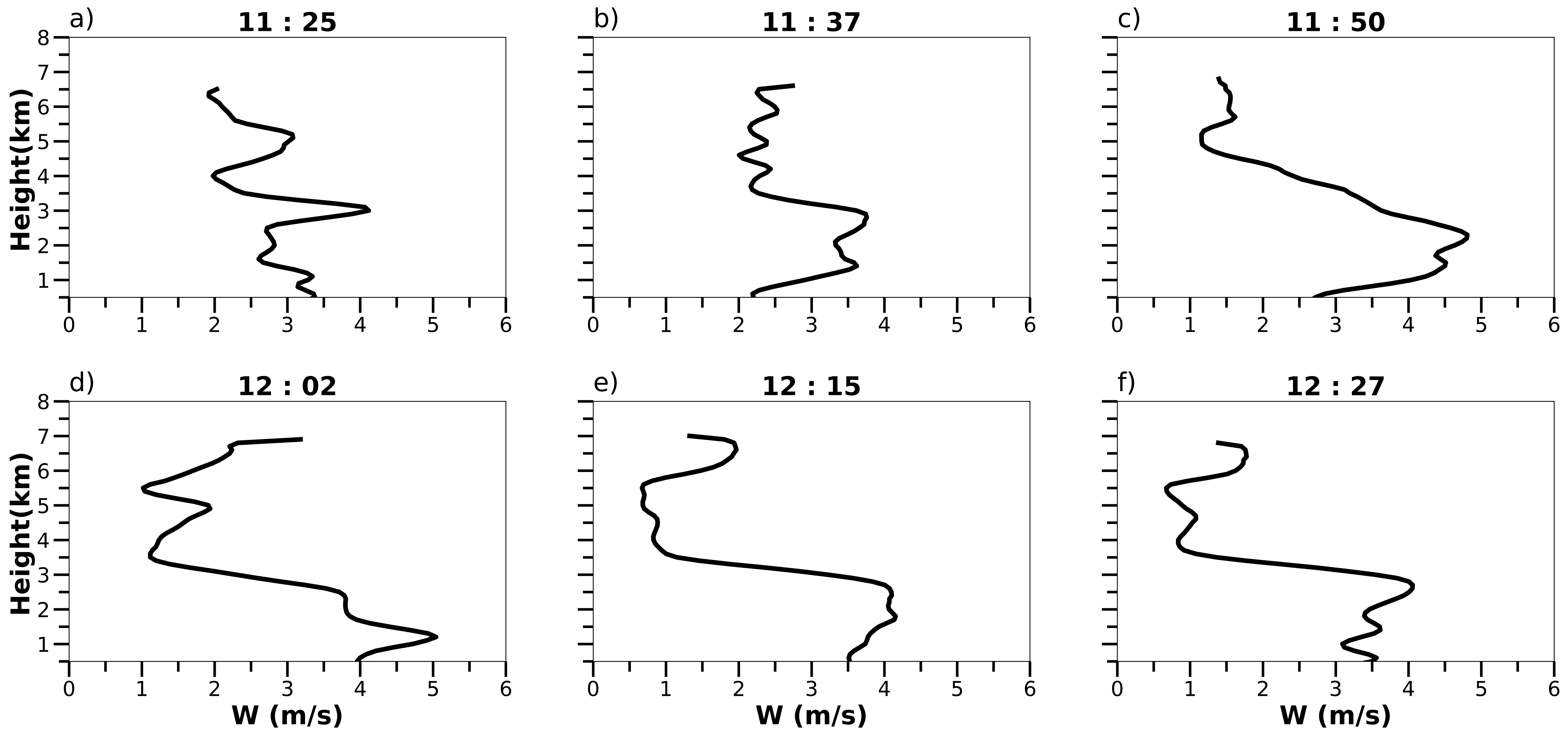}
    \textbf{The vertical profiles of Z(FIG 7), DV(FIG 8) and W(FIG 9) are analysed every \~12 minutes using PPI scan data taken at 85º elevation(pseudo zenith looking). All times given in UTC}

    \label{FIG 9}
\end{figure}

\begin{figure}[hbt!]
    \centering
    \includegraphics[width=0.75\linewidth]{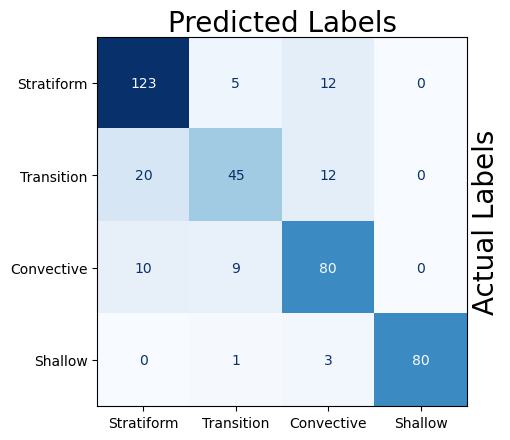}
    \renewcommand{\thefigure}{10}
    \caption{\textbf{ Confusion Matrix. The main diagonal elements represent the TPs of each type of cloud. Excluding the TPs, the sum of row elements for each cloud indicates FNs and the sum of column elements indicates FPs. The sum of every remaining element indicates TNs for a single type of cloud}
}
    \label{FIG 10}
\end{figure}
\begin{figure}[hbt!]
    \centering
    \includegraphics[width=1\linewidth]{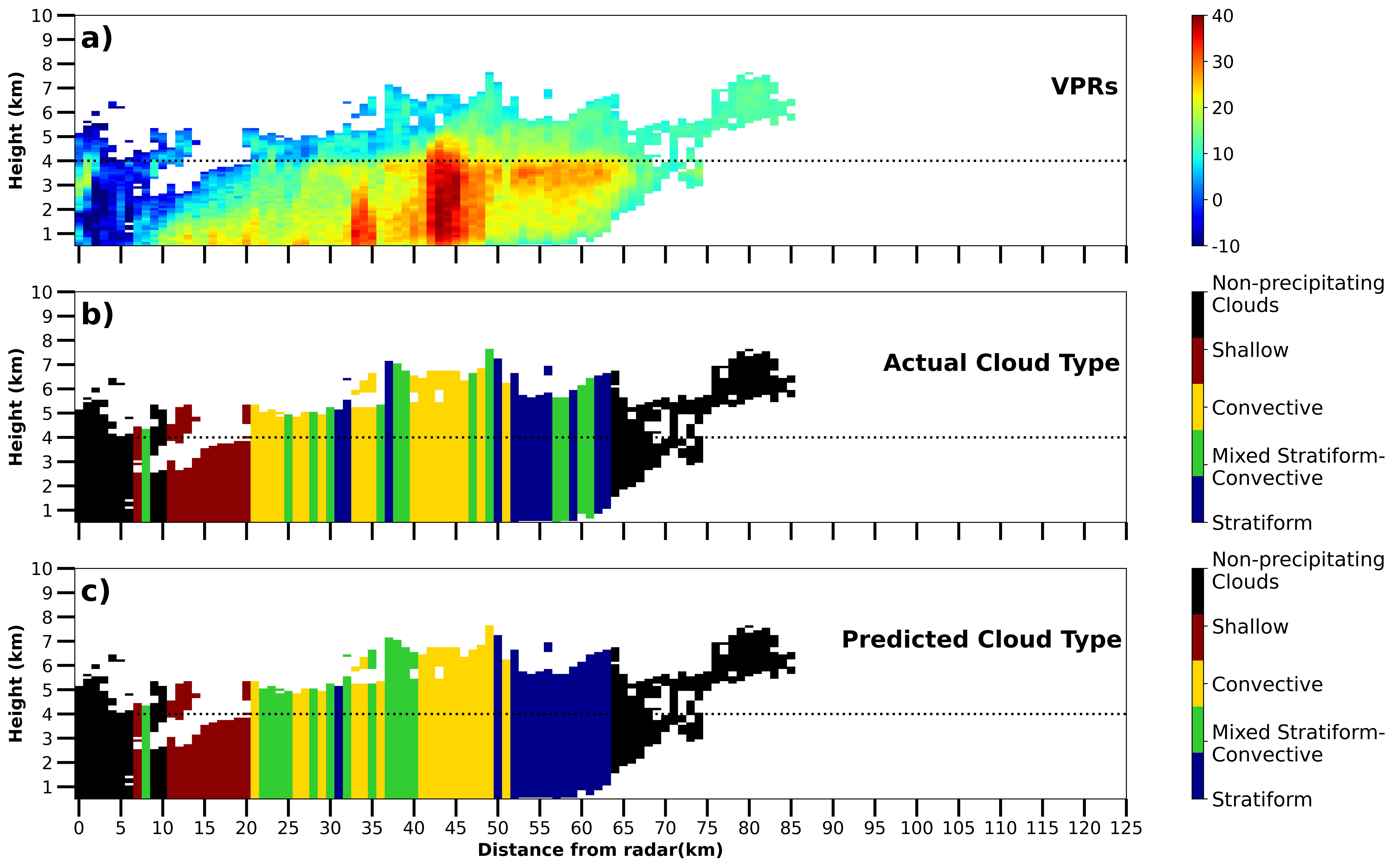}
        \renewcommand{\thefigure}{11}
    \caption{\textbf{ a) RHI scan taken by the radar at 3:52 UTC on July 15 2018 b) Corresponding Actual/labelled cloud profiles and c) Predicted cloud profiles by the model. The dotted line at 4 km indicates the melting layer.}
}
    \label{FIG 11}
\end{figure}
\begin{figure}[hbt!]
    \centering
    \includegraphics[width=0.75\linewidth]{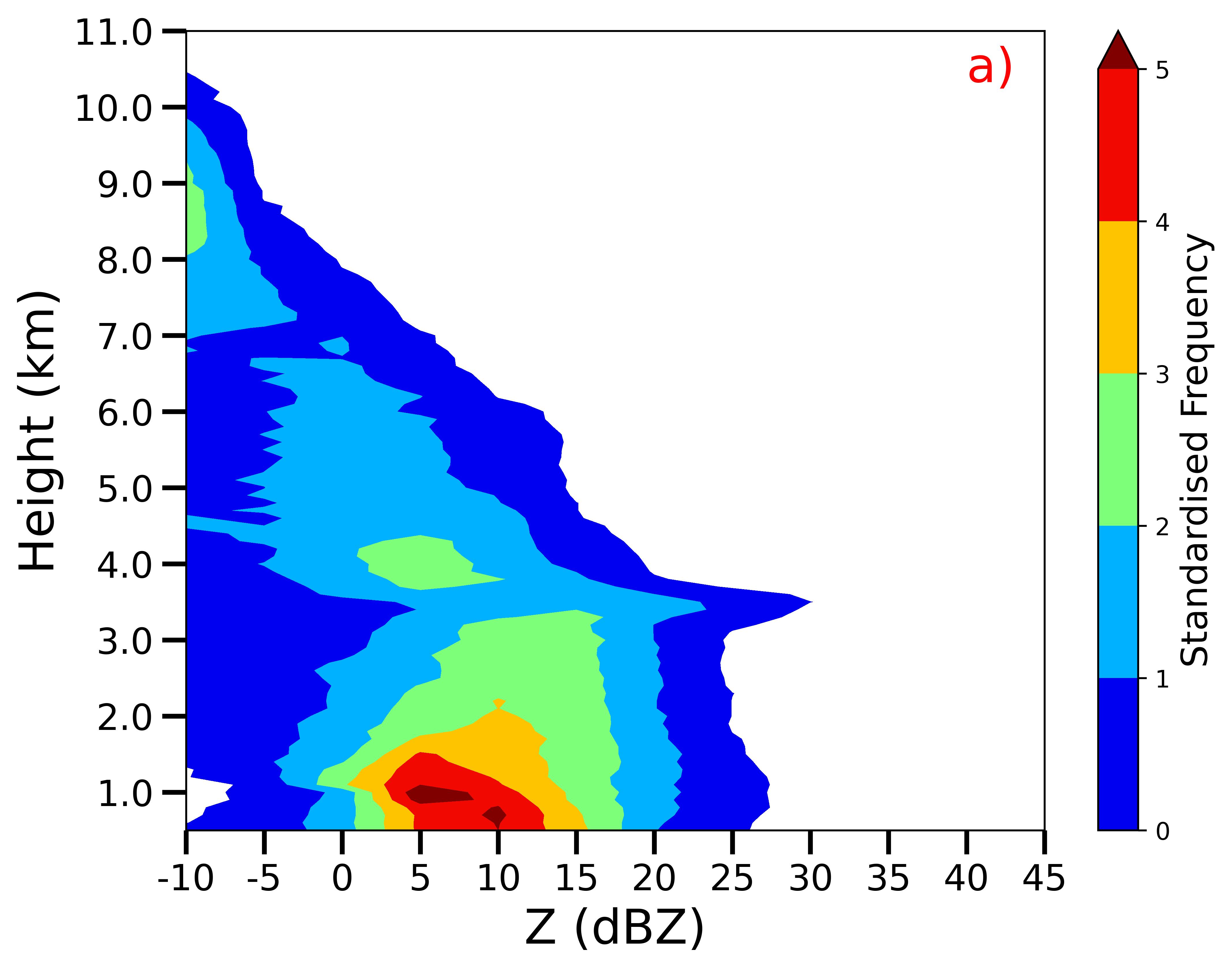}
\end{figure}
\begin{figure}[hbt!]
    \centering
    \includegraphics[width=0.75\linewidth]{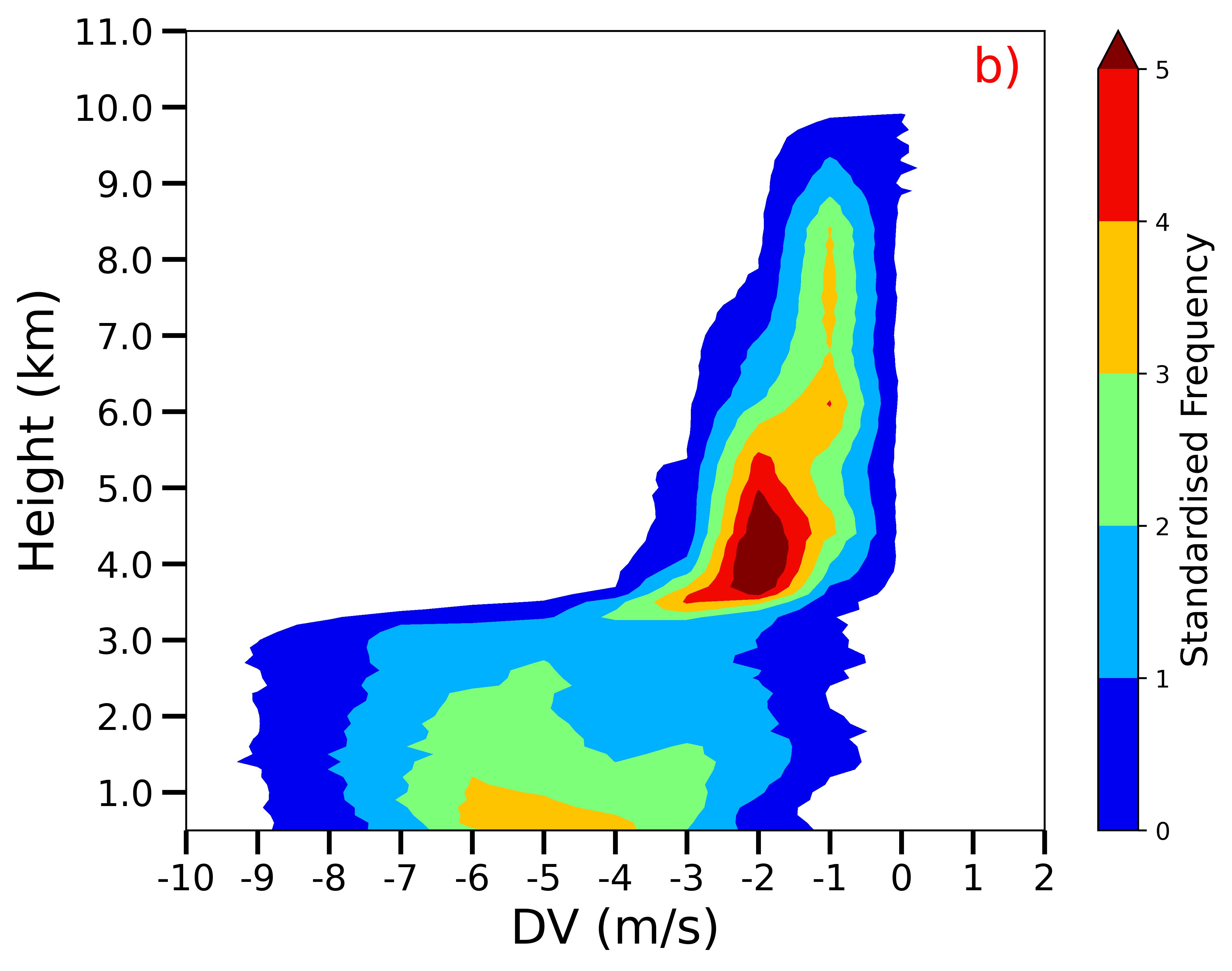}
\end{figure}
\begin{figure}[hbt!]
    \centering
        \renewcommand{\thefigure}{12}
    \includegraphics[width=0.75\linewidth]{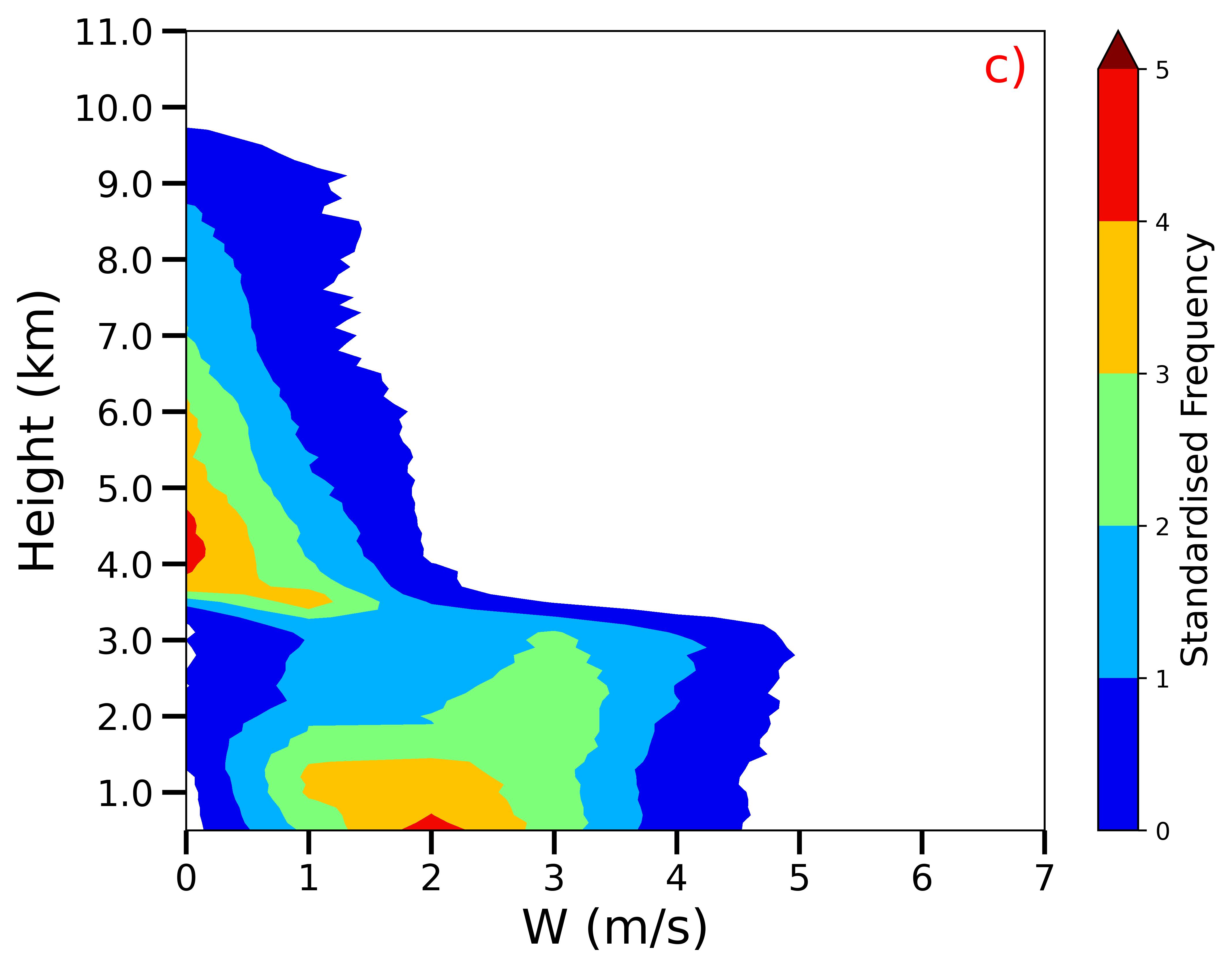}
    \caption{\textbf{ CFADs of a) Z b) DV c) W constructed over the radar site (MDV) site}
}
    \label{FIG 12}
\end{figure}


\end{document}